# Neo-humanism and COVID-19:
# Opportunities for a socially and environmentally sustainable world

Francesco Sarracino and Kelsey J. O'Connor

Version: 20 April 2021

A series of crises, culminating with COVID-19, shows that going "Beyond GDP" is urgently necessary. Social and environmental degradation are consequences of emphasizing GDP as a measure of progress. This degradation created the conditions for the COVID-19 pandemic and limited the efficacy of counter-measures. Additionally, rich countries did not fare the pandemic much better than poor ones. COVID-19 thrived on inequalities and a lack of cooperation. In this article we leverage on defensive growth models to explain the complex relationships between these factors, and we put forward the idea of neo-humanism, a cultural movement grounded on evidence from quality-of-life studies. The movement proposes a new culture leading towards a socially and environmentally sustainable future. Specifically, neo-humanism suggests that prioritizing well-being by, for instance, promoting social relations, would benefit the environment, enable collective action to address public issues, which in turn positively affects productivity and health, among other behavioral outcomes, and thereby instills a virtuous cycle. Arguably, such a society would have been better endowed to cope with COVID-19, and possibly even prevented the pandemic. Neo-humanism proposes a world in which the well-being of people comes before the well-being of markets, in which promoting cooperation and social relations represents the starting point for better lives, and a peaceful and respectful coexistence with other species on Earth.

Sarracino (corresponding author): Senior Economist, Institut national de la statistique et des études économiques du Grand-Duché du Luxembourg (STATEC Research). GLO Fellow, Global Labor Organization. 14, rue Erasme, L-2013, Luxembourg. +352 247 88469. Francesco.Sarracino@statec.etat.lu.

O'Connor: Economics Researcher, Institut national de la statistique et des études économiques du Grand-Duché du Luxembourg (STATEC Research). GLO Fellow, Global Labor Organization (GLO). Research Affiliate, Institute of Labor Economics (IZA). Kelsey.OConnor@statec.etat.lu.

**Acknowledgements:** The authors would like to thank the participants at the International Society for Quality-of-Life Studies Annual Conference 2020 for valuable comments, and gratefully acknowledge the financial support of the Observatoire de la Compétitivité, Ministère de l'Economie, DG Compétitivité, Luxembourg, and STATEC. This article reflects the views of the authors and does not engage in any way STATEC, STATEC Research or funding partners. This research did not receive any specific grant from funding agencies in the public, commercial, or not-for-profit sectors.

# 1. Introduction

Neo-humanism is a cultural movement to put humankind back at the center of decision making (Sarracino 2020). Traditional economic thinking elevated GDP per capita to the single-most important indicator of quality of life, used explicitly by policy makers and implicitly by civil society. We argue the emphasis on income has not served us well in recent years, generally, and in particular during the COVID-19 pandemic; it contributed to the degradation of both the natural and social environment, which, in turn, increased the chances of a pandemic, and limited the efficacy of countermeasures. The origins of COVID-19 can probably be traced back to environmental degradation, while the emphasis on individualism limited our ability to leverage social capital to effectively coordinate a response. Neo-humanism invites us to expand our focus, from the singular dimension of economic output towards a more holistic concept of quality of life. Quality-of-life studies have gone a long way to inform neo-humanism, and social capital -- the cultural fabric that allows a society to cooperate to achieve common goals -- plays a central role. It is time to distill and disseminate this knowledge to create a new culture leading towards a socially and environmentally sustainable future.

The pursuit of economic growth at any cost plausibly weakened countries' ability to cope with COVID-19. Countries with greater GDP per capita did not perform better, while the emphasis on economic growth plausibly diminished social capital and limited the efficacy of countermeasures to COVID-19. What is more, the initial transmission of COVID-19 to humans is likely partially due to environmental degradation, which in turn, is a consequence of mismanaged growth.

Defensive growth theory helps us making sense of this puzzle. The theory describes the negative interactions between growth and well-being, with specific implications pertaining to social capital and the environment. We know that economic growth occurs with a rise in demand, including when non-market public resources – such as a pristine environment – are substituted with private goods, e.g. private yards and entertainment equipment. Such growth, arising from the substitution of private goods for diminished relational and public goods -- referred to as "defensive growth" -- creates, and accrues from, a vicious cycle whereby the additional degradation of public resources, fuels further consumption of private goods, in a self-reinforcing loop. Defensive growth models provide an explanation for certain paradoxical facets of modern society: long working hours; emphasis on consumption and material concerns; unhappiness; decreasing social capital; and environmental degradation. Defensive growth theory also provides an explanation of why modern societies are far from sustainable. According to this theory, unsustainability originates from the organization of modern society, not from human greed. The implication is that the key to environmental sustainability and to quality of life is re-orienting social and economic activities to prioritize people over markets. This, in turn, means abandoning the myth that well-functioning markets strictly lead to better lives.

We leverage on insights from the quality-of-life literature to argue that it is possible to promote a virtuous cycle in which investing in social relations and well-being reduces people's need to consume, thus protecting the environment and promoting social relationships. Indeed, greater well-being leads to efficiency gains which can be used to reduce working time and ultimately decouple well-being from defensive, or palliative, consumption.[1] We conclude that it is possible to organize modern societies according to a virtuous cycle in which the explicit pursuit of well-being through policies, such as those promoting social capital, contributes to a socially and environmentally compatible economic growth.

In what follows we discuss how environmental degradation increased the risks of pandemics to occur, like COVID-19. In Section 3 we discuss the impacts of COVID-19, while in Section 4, we discuss the differential impact of COVID-19 across countries. Section 5 pertains to defensive growth theory, which describes a vicious cycle in which the economic bads (negative externalities) generated by growth contribute to more economic growth. In Section 6, we describe neo-humanism and how it could lead to a reorganization of society that puts quality of life before economic growth. The last section concludes.

## 2. Origins of COVID-19: Environmental Degradation

A number of researchers agree that environmental degradation, in particular the loss of biodiversity, create the conditions for new viruses and infections, like COVID-19, to spread. Undisturbed ecosystems operate in a delicate balance, which if upset, can lead to the proliferation of pests (Barouki et al., 2021; Vidal, 2020; AllEnvi, 2020; Sigal, 2020). Biodiversity is another way to think of this balance. It can be conceived of as a barrier that keeps naturally occurring pathogens in balance and away from humans. The loss of biodiversity increases the chances that humans become exposed to various pathogens. Lyme disease serves as an example.

---

[1] Defensive consumption is palliative, that is, it provides temporary relief (defense) from the degradation of non-market resources, but does not address the problem.

Lyme disease was first detected in 1975 in the town of Lyme in Connecticut -- Northeast coast of the United States. The disease is caused by a bacterium transmitted by the bite of blacklegged tick. The infection can cause skin rash, fever, headache, fatigue and, if untreated, can have serious health consequences for joints, the heart, and the nervous system. The bacterium has always existed – as documented in various chronicles, but the number of infected had remained small. So, what changed in in the town of Lyme leading up to 1975?

The northeast coast of United States used to host a rich and flourishing forest characterized by numerous plant and animal species. However, the forest has been undergoing a long-term process of deforestation due to logging and the expansion of towns and suburbs. By damaging the ecosystem, and reducing its diversity, many of the species that inhabited the forest disappeared. Among these were opossums and chipmunks, two formidable predators of ticks. In absence of a natural predator, the number of blacklegged ticks rose. This, along with the expansion of towns and suburbs towards the forest, created the conditions for "a perfect storm": reducing the distance between humans and a large population of ticks, the probability that the disease passed onto humans grew greatly. However, there is a big difference between Lyme disease and COVID-19: both originated from animals, but COVID-19 is transmitted by humans, whereas Lyme disease needs a vector, the tick, to reach humans; that is why Lyme disease never turned into a pandemic.

The explanation for the rise of the Lyme disease can be applied to the emergence of other infectious diseases. A growing body of environmental research shows that over the last 40 years the number and diversity of outbreaks, and richness of diseases increased significantly. The upper left bar plot in Figure 1 shows the cumulative number of outbreaks over time, along with the number of events (richness) constituting each outbreak. Figure 1b shows that nearly half of these new infections are of zoonotic origins, that is they are due to contagions from wild or domestic animals, as is the case for COVID-19. These data suggest that infections such as COVID-19, the swine flu, SARS or Ebola, represent only well-known diseases that eventually reached the news, but in fact there are many more infectious disease outbreaks occurring each year. The frequency of new infections has increased over time, and more and more infections are caused by viruses and bacteria (Figure 1c). The problem is, when the number of outbreaks increases, so does the probability that one of these outbreaks turns into a pandemic.

*Figure 1. Global number of human infectious disease outbreaks and richness of causal diseases 1980–2010.*

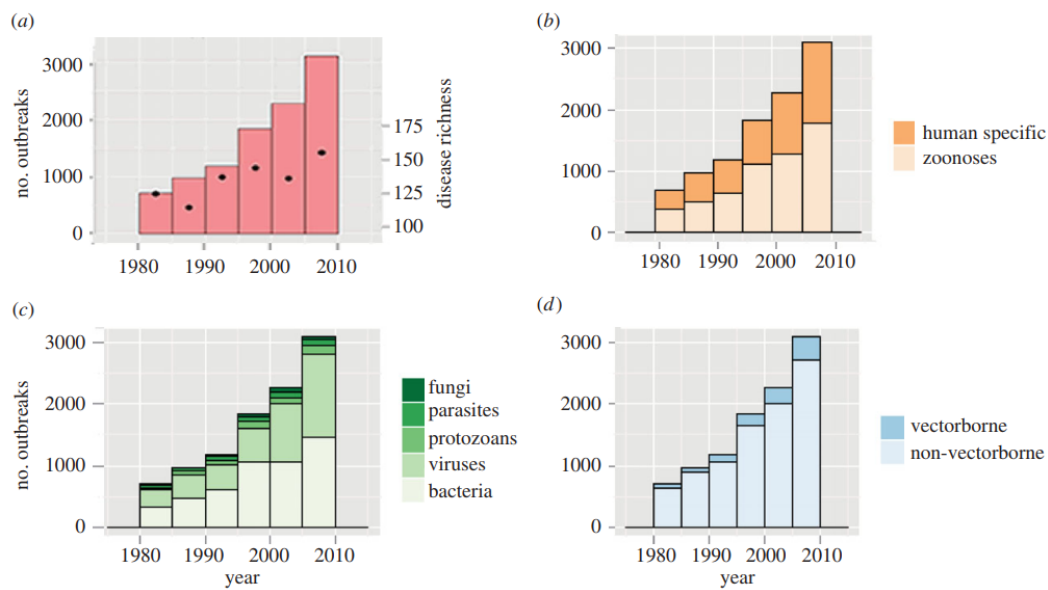

*Source: Smith et al., 2014, p. 2.*
*Note: Outbreak records are plotted with respect to (a) total global outbreaks (left axis, bars) and total number of diseases causing outbreaks in each year (right axis, dots), (b) host type, (c) pathogen taxonomy and (d) transmission mode.*

The increasing number of outbreaks is consistent with the evidence that biodiversity has decreased over time. Biodiversity has declined at an accelerating rate over the past 100 years; this is observed in the form of increasing extinctions in Figure 2. Ceballos and colleagues (2015) compared these estimates with the estimated background rate of extinction. The background rate represents the rate of extinction absent human activities, typically estimated from the fossil record. In order to respond to the skeptics of human-induced species loss, the authors spend a significant amount of time and apply highly conservative assumptions to estimate the background rate. Even based on their highly conservative estimates, extinctions are an order of magnitude above the expected background rate of extinctions. The authors attribute causes to human population size and growth, which in turn affects consumption (especially in rich countries), habitat loss, and climate change.

Thus, human action has almost certainly facilitated the emergence and spread of COVID-19. The evidence suggests human population growth and consumption contributed to the loss of biodiversity, which in turn threw delicate ecosystems out of balance, reducing biodiversity and giving rise to the conditions to increase the number of pathogens. Humans also increased their exposure by moving more and more into previously relatively undisturbed habitats around the world. As it happened in other well-known cases of infectious diseases, human action likely increased the number of Corona viruses and risk of exposure.

*Figure 2. Cumulative vertebrate species recorded as extinct or extinct in the wild by the International Union of Conservation of Nature (2012).*

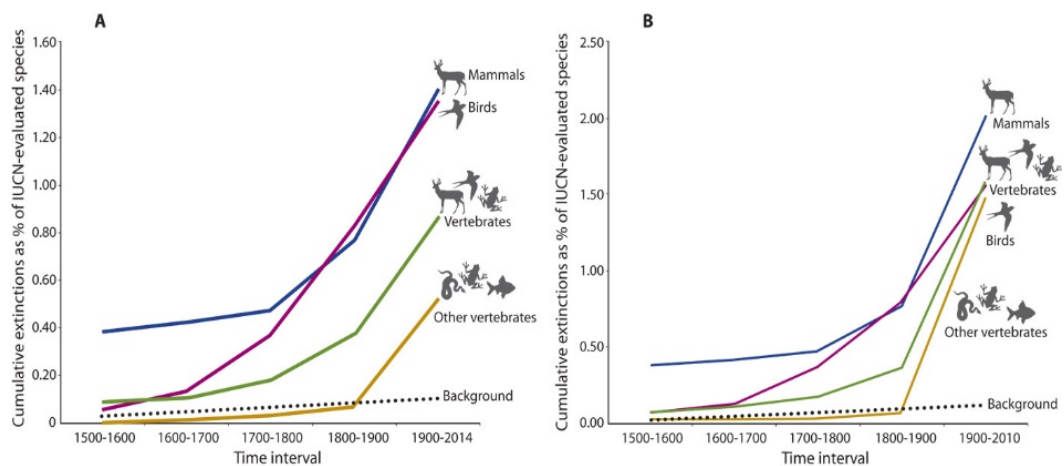

*Note 1. Graphs show the percentage of the number of the number of species evaluated among mammals (5513; 100% of those described), birds (10,425; 100%), reptiles (4414; 44%), amphibians (6414; 88%), fishes (12,457; 38%), and all vertebrates combined (39,223; 59%). Dashed black curve represents the number of extinctions expected under a constant standard background rate. (A) Highly conservative estimate. (B) Conservative estimate*
*Source: Ceballos et al., 2015, p.3.*

This was well-known to American intelligence experts: in the "Worldwide Threat Assessment" of January 2019, the U.S. Intelligence Community raised insistent concerns about the risks of a pandemic. The report reads: "we anticipate [there] will be more frequent outbreaks of infectious diseases because of rapid unplanned urbanization, prolonged humanitarian crises, human incursion into previously unsettled land, expansion of international travel and trade, and regional climate change" (Coates 2019, p. 21). In sum, it is not bats or pangolins, per se, that pose a threat to public health. The threat comes rather from human action.

## 3. COVID-19 Impacts

It is not possible to enumerate the great many consequences of the COVID-19 pandemic around the world. Only the most obvious impacts are on physical health and the economy. By the end of February 2021, nearly 2.5 million people (i.e. about 316 people per million) died because of COVID-19 worldwide. This, however, only captures the reported deaths due directly to COVID-19. Additional deaths occurred due to strain on health infrastructure and access. Data from EuroMOMO[2], a network of epidemiologists who collect data on all-cause mortality in 24 European countries, indicate that excess mortality due to COVID-19 has been far greater in 2020 than in the preceding 10 years. Excess mortality is the number of people who die from any cause in a given region and period compared to a historical baseline. Figure 3 presents the figures by country. All but Norway in the sample experienced increased excess mortality.

*Figure 3. Excess All-Cause mortality (per 100,000 people) in a selected group of countries.*

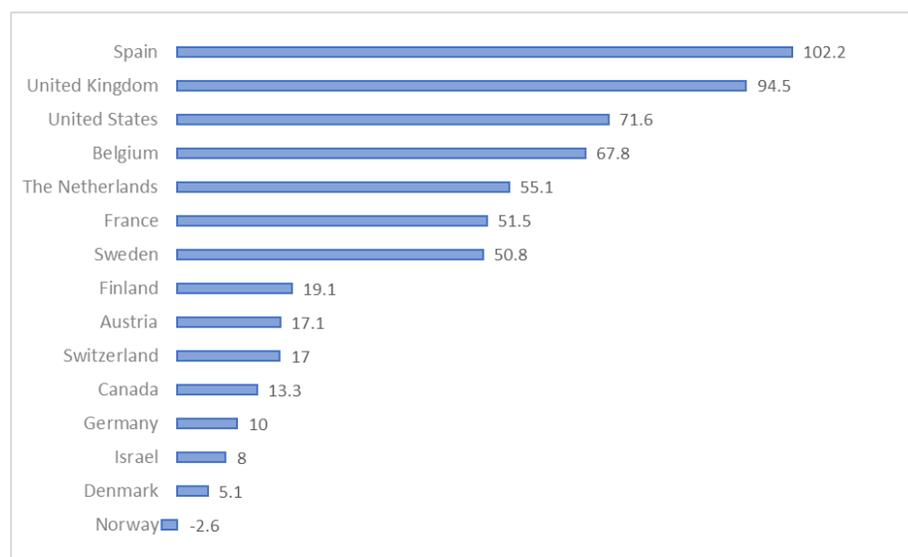

*Note: Data on deaths are through July 25, 2020. Countries lacking publicly available all-cause mortality data through this time are omitted. Excess deaths were estimated by week, compared with 2015-2019, beginning when a country surpassed 1 COVID-19 case per million population.*
*Source: Bilinski and Emanuel, 2020.*

The impacts have not been felt equally. Available figures suggest that COVID-19 worsens existing inequalities and probably contributes to new inequalities along unprecedented dimensions. For instance, a recent study by the statistical office of Great Britain (ONS), shows that black people, Bangladeshi and Pakistani have nearly two times higher chances of dying from COVID-19 than whites (see Figure 4). Pre-existing health conditions, such as diabetes, asthma, hypertension, kidney disease, and obesity contribute to differential rates. Additionally, certain jobs and lives of people are riskier than others. Vulnerable people live in more crowded neighborhoods, in smaller houses, experience greater income volatility, and frequently, their jobs cannot be performed remotely (Adams-Prassl et al., 2020).

---

[2] https://www.euromomo.eu/.

Those who earn less money are less able to protect themselves from infection. Adams-Prassl and colleagues administered a survey on two samples of American and British residents, finding that workers earning more than $70,000 per year can perform more than 60% of their work tasks from home, whereas the corresponding figure for those earning less than $40,000 is less than 40%. Additional evidence comes from Google Mobility Data in United States, which reveal that people living in the richest 10% of counties reduced their travel by 39%, while those in the poorest 10% cut their movements by 27% (The Economist, 2020).

The same divide holds by education. For instance, in Luxembourg a higher education degree is associated with greater opportunities to work from home (see Figure 5): more that 69% of people with a master degree or higher could work remotely, whereas this was possible for less than 25% of people with a lower secondary or primary education.

*Figure 4. Ethnic minorities in England have higher chances to die because of COVID-19 than White.*

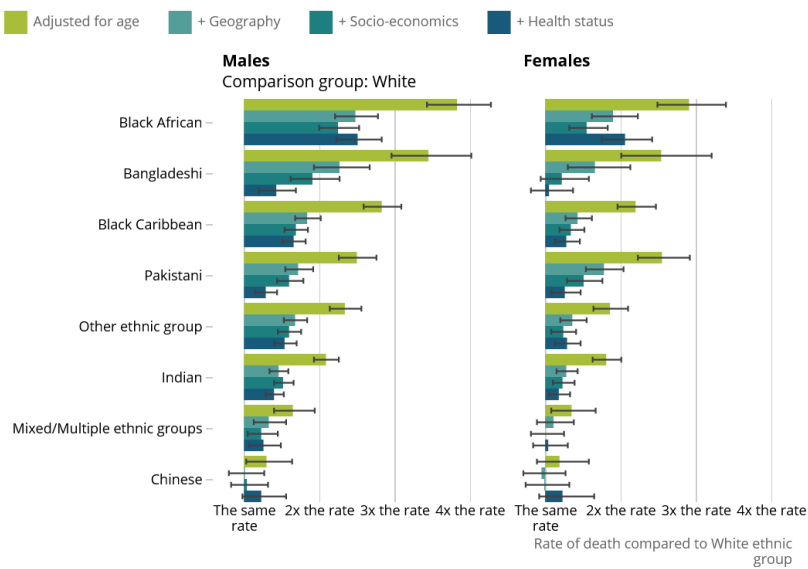

*Source: Office for National Statistics. Explaining ethnic background contrasts in deaths involving Coronavirus (COVID-19).*
*Note: Rate of death involving COVID-19 by ethnic group and sex relative to the White population, England, 2 March to 28 July 2020.*

*Figure 5. Possibility to work remotely by education in Luxembourg.*

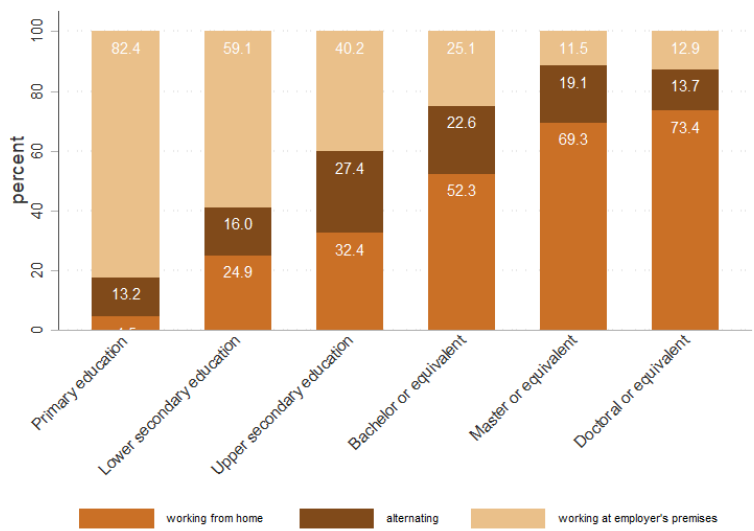

COVID-19 also created new inequalities. While many became familiar with a booming number of web applications to communicate with others, not everyone had access to fast internet connections, powerful PCs or smartphones. Nor does everyone have the technical skills necessary to use these technologies, for instance: digital analphabets, elderly, poorer or less educated people who, for instance, are not familiar with the use of these technologies, simply cannot afford them, or must share one computer and one connection with the whole family. These tools have often become necessary to conduct important activities remotely, school work, professional activities and to stay in contact with loved ones, with family members, friends, and colleagues. Limited access or understanding, therefore, became important new sources of inequality and exclusion, posing significant challenges to society. These are particularly salient for young people. For instance, students learn to socialize in schools in face-to-face relationships. COVID-19 forced them (those with access) to web-mediated relationships more than before, and at an even earlier stage in life than before. It is unclear what socio-economic consequences this may have.

In sum, the pandemic had far-reaching effects, not the least of which are existing and new inequalities. This will have unexpected and unpredictable impacts on people's well-being. For instance, previously the most vulnerable to social isolation were those outside the job market. Now the risk of social isolation has exploded generally and along dimensions that are generational, ethnic, income-related, regional, educational, and related to the family of origin. Social isolation and loneliness are general health risks (mental and physical) (Luo et al. 2012), while inequalities reduce perceived fairness and trust (Oishi et al. 2011), both necessary for social cohesion and cooperation. There are also reasons to be concerned about the implications of such changes for the future.

## 4. Country response to COVID-19

Some countries fared the pandemic better than others. As shown earlier in Figure 3, excess mortality differs considerably across countries. Figure 6 yields similar observations. The Case Fatality Rate (i.e., deaths per 100 positive cases) also differs considerably. It is too early to answer conclusively why some countries fared better than others. Indeed, the answer depends on the metric and likely involves multiple facets. We do know, however, that countries adopted different sets of measures at different points in time. Physical distancing, tracking of positive cases, and lockdowns are some of the most widely adopted measures to "flatten the curve" (of infections) (OECD, 2020).

A considerable amount of research has evaluated the effectiveness of containment policies to limit the contagion. For instance, results indicate that physical distancing (typically imposed by lockdown) worked as expected (O'Connor 2020a). Figure 7 reports the relationship between the day when increased physical distancing occurred (as measured on the x-axis) and the time to reach the peak in new infections (on the y-axis). The scatterplot indicates that countries which more quickly responded to the first positive case in their country (with significant distancing), reached the peak in new infections earlier, thereby reducing the severity of the contagion.

Countries differed markedly in the timing and extent of lockdown measures, presumably in large part because of their heavy social and economic costs, costs which many question the value of. This position is exemplified by Mr. Donald Trump who, in March 2020, tweeted: "we cannot let the cure be worse than the problem itself." Nearly one year later, we better understand the significance of the problem. Given the number of victims, variants of the virus, and an unforeseeable end, stronger treatment would have been a significant improvement.

*Figure 6. COVID-19 Case Fatality Rate across countries (Feb. 21, 2021).*

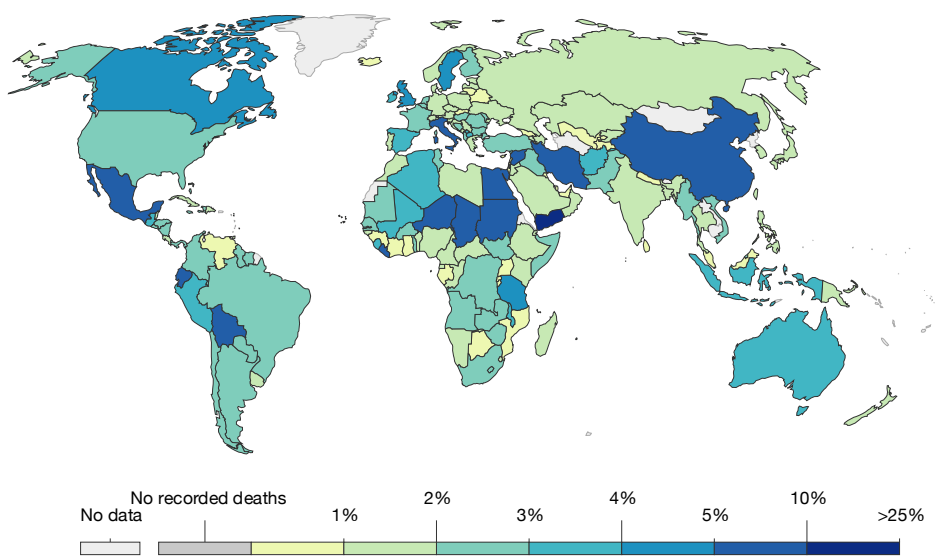

*Source: John Hopkins University CSSE COVID-19 Data - Last updated 22 February, 09:03 (London time). Accessed via Our World in Data.*

*Note: Case Fatality Rate is the share of confirmed deaths over positive cases of COVID-19.*

*Figure 7. Countries that introduced the lockdown later, reached the peak of new infections later.*

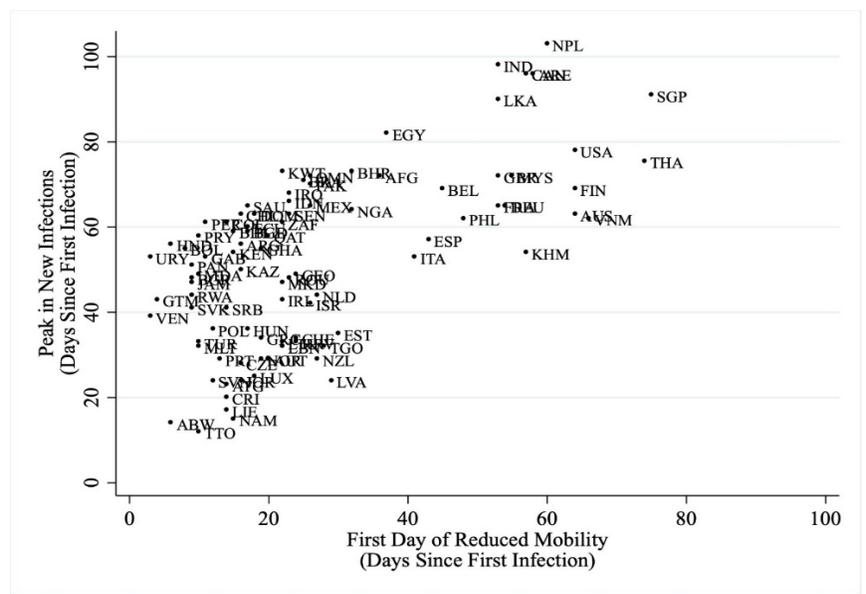

*Source:* (O'Connor 2020a)

*Note: Data on new infections are retrieved from OurWorldinData.org. Mobility restrictions are issued from Google Mobility Data. The data refer to the first wave of the pandemic.*

Mr. Trump was not alone: many others believed and still think that the economic costs of lockdown are too great, which shows how much the equation economic growth equals better lives is endemic in modern culture. If the economy is a tool created by mankind to better organize their life, there should be no conflict between protecting lives and the economy. Yet, that is what happened in many countries. COVID-19, just like the economic crisis of 2008, exposes the limits of this economy-first culture and illustrates how urgently we should reform modern societies.

In fact, we expect rich countries to be better prepared to face the pandemic than poor ones, as they can rely on better infrastructure, mass screening, and medical knowledge and technologies to limit fatality rates of the virus. And this is arguably undisputed. However, the scatterplot in Figure 8 plots countries' wealth (as measured by GDP per capita in 2018) against Case Fatality Rate: richer countries do perform better, but only weakly. The results are even more discouraging when accounting for confounding variables. From the same study behind Figure 7, regression results show countries with greater Gross National Income per capita experienced more severe first waves (O'Connor 2020a).[3] Additionally, Deaton (2021) documents a negative correlation between mortality and (log of) per capita income. This evidence is unsettling and questions one of the cultural pillars of modern countries: the belief that growing economies are the gateway to better lives.

---

[3] Countries with greater GNI per capita reached the peak in new infections later (although insignificantly), and experienced a greater number of new infections per day when they reached the peak (significantly). The regressions included the additional explanatory variables: distancing behavior, population density, the population share that is 65 years or older, total population, the capital city's latitude, an index of global interconnections, an index of democracy, and the average number of years of school.

*Figure 8. There is little association between Case Fatality Rate and GDP per capita worldwide.*

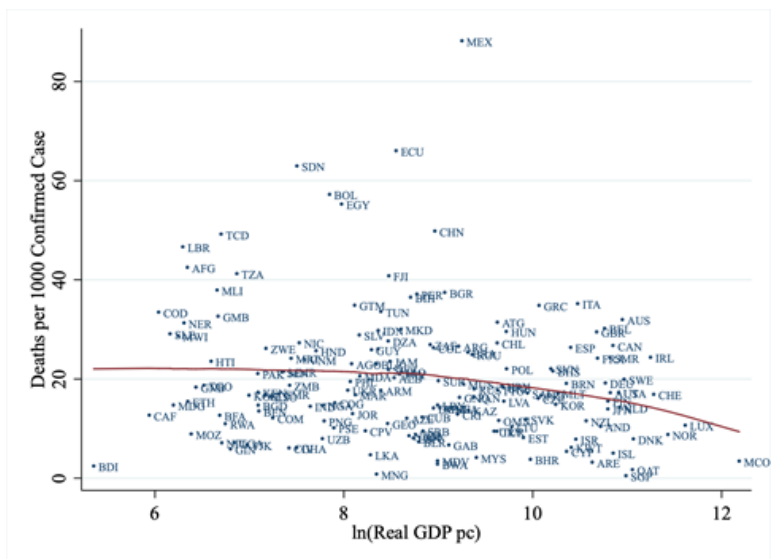

*Source: own elaboration of John Hopkins University and Medicine data and World Development Indicators. Data cover the year 2020. On the y-axis we report mortality per 1000 confirmed cases. The x-axis orders countries by gross domestic product per capita, in real international dollars of 2010 and adjusted using a logarithmic function.*

How is it possible that rich countries, with access to better technology and know-how, did not perform significantly better than poorer ones? Part of the explanation may be due to under-reporting in poor countries. However, the association between GDP per capita and fatality rates is even weaker if we consider only the set of member States of the Organization for Economic Co-operation and Development (OECD) -- a group where measurement errors should be less of an issue. Descriptive statistics suggest that rich countries could have fared the crisis better than others if they had been better prepared and more cooperative.

## 4.1 Unprepared and uncoordinated

Rich countries were unprepared and their responses uncoordinated. Countries around the world restricted travel in an effort to reduce the transmission of COVID-19. Countries also competed with each other over scarce medical resources. It should be relatively clear that the best way to combat a global pandemic is to invest scarce resources where they have the largest impact, and as infection rates evolve non-linearly, coordination becomes particularly relevant. Few cases in one province of one country exploded around the world in less than half a year.

Rich countries acted late on average. Based on Google Mobility Data, Figure 9 presents the distribution of the number of days (since the first positive case) necessary to increase physical distancing[4] by 30% or more. The plot on the right indicates that the richest 50% of OECD countries took longer on average to introduce lockdown than those belonging to the bottom 50%[5]. Also, the size of the box shows that rich countries had a very heterogeneous approach to lockdown. This conclusion is also supported by data on response stringency provided by the University of Oxford (Hale et al, 2020).

The European Union probably would have faced the pandemic more effectively if countries had better coordinated their responses. EU coordination is particularly important given common goods and labor markets make it more difficult to restrict flows across borders. The Oxford Stringency Index, which summarizes the policies adopted to contain the contagion, indicates policies were not well coordinated. Figure 10 presents its distribution seven days after the first positive case and indicates large disparities in the relatively homogeneous group of countries. On a scale from 0 to 100, the index ranges from near-zero (Estonia, Sweden or The Netherlands) to well above fifty (Hungary, Bulgaria, Slovakia). As countries did not experience the infection at the same time, differences in adopted policies are understandable. However, seven days after the first positive case, countries could have adopted common strategies, coordinated at European level. This did not happen and it favored the proliferation of viral variants.

---

[4] Physical distancing is derived from Google mobility data as the negative of the average amount of time spent in the location types: retail and recreation, parks, workplaces, grocery and pharmacy, and transit stations (O'Connor 2020a). Time spent is reported relative to a baseline period prior to the pandemic. Thus, an increase in distancing means less time spent in these locations relative to the baseline period.

[5] The countries below the median of Gross National Income per capita are Czech Republic, Estonia, Greece, Hungary, Latvia, Lithuania, Portugal, Slovakia, Slovenia, and Spain. Those above are: Austria, Belgium, Finland, France, Germany, Ireland, Italy, Luxembourg, The Netherlands, and United Kingdom. We use Gross National Income, rather than GDP, because of the presence of some small countries, such as Luxembourg, for which GNI is a better measure of the wealth of a country. However, GNI and GDP are strongly associated, thus our results are not sensitive to this choice.

*Figure 9. Rich countries acted late.*

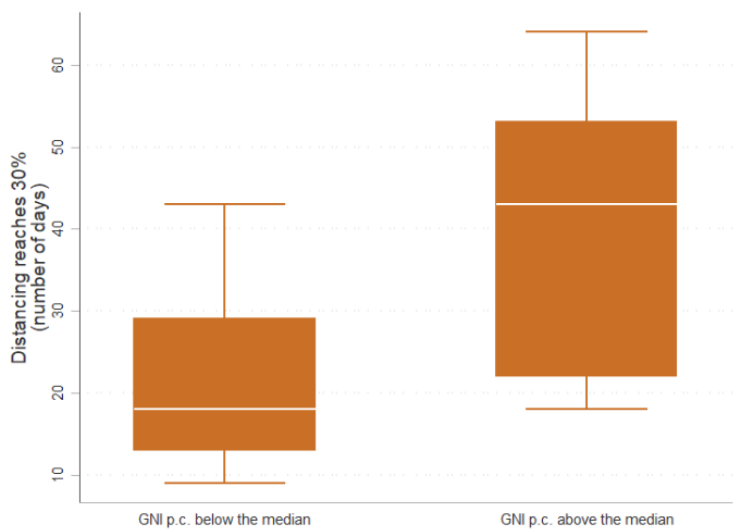

*Source: Own elaboration of Google Mobility Data (May 2020) and World Development Indicators (2018).*

*Figure 10. Rich countries did not act together (24 European Union countries). The response stringency index ranges from 0 to 100, where higher scores indicate more stringent policies.*

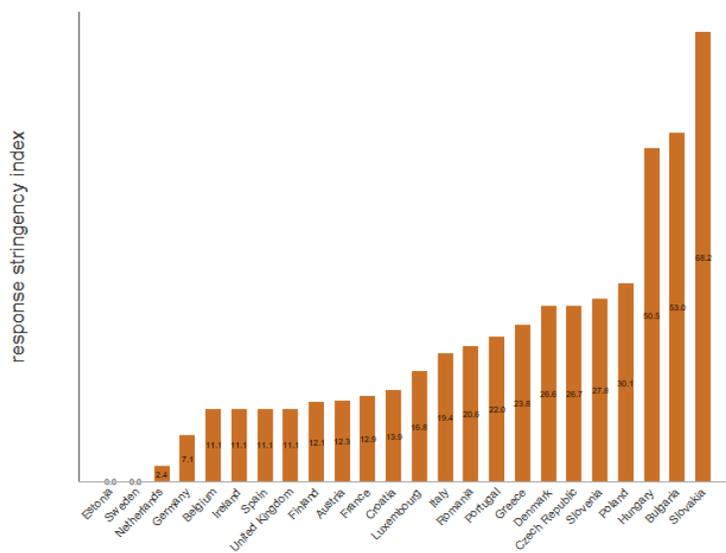

*Source: own elaboration of Our World in Data (May 2020).*

Rich countries could have been better prepared. Anecdotal evidence indicates that both European Union and United States did little to prepare for epidemics despite repeated warnings from experts. For instance, the previously-mentioned "Worldwide Threat Assessment" report of the U.S. Intelligence Community stated, "We assess that the United States and the world will remain vulnerable to the next flu pandemic or large scale outbreak of a contagious disease that could lead to massive rates of death and disability, severely affect the world economy, strain international resources, and increase calls on the United States for support" (Coates 2019, p. 21). One year later, the US was little prepared to face COVID-19, and Europe was not very different. Ms. Von der Leyen, the President of the European Union, announced the creation of a stockpile of medical equipment for the European Union on the 19th of March, 2020 – when some European Countries were already about to reach the peak in new contagions.

Another example of how unprepared the rich countries were, is the time elapsed before they could administer a significant number of COVID-19 tests. Figure 11 presents the number of tests administered per 1000 people after seven, 14, and 40 days from the first infection in the country: the majority of countries took more than 40 days to administer at least 10 tests per 1000 people. Despite some exceptions, such as Iceland and Luxembourg, the majority of countries were not able to implement a significant testing campaign in a timely manner. In sum, rich countries were unprepared for and poorly coordinated their response to COVID-19, which partially explains why they have not fared better than others.

*Figure 11. Number of tests administered on the general population at various points in time after the first positive case. The countries included are OECD member states.*

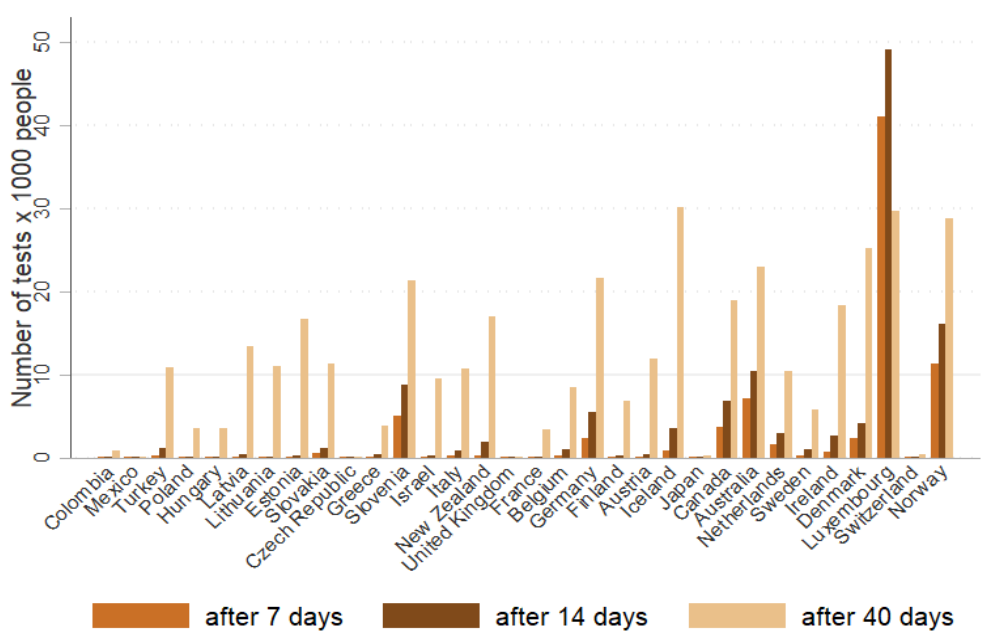

The values for Iceland and Luxembourg after 40 days have been divided by 3 for graphic representation.

*Source: own elaboration of John Hopkins University and Medicine data (May 2020).*

## 4.2 The role of social capital

The effectiveness of countermeasures to fight epidemics depends largely on human behavior, in particular collective action, in which everyone's actions matter: from those who are committed on the front line of the fight to contain and treat infection, to those who patiently wait at home and respect containment policies. Compliance with containment policy is ripe for freeriding: it is costly for individuals to comply with containment policies, especially when it could affect their employment. If individuals ignored social costs and responded solely to individual incentives, then few would comply and containment policies would fail. Cooperation is necessary for containment policy to work. That is why social capital is one of the important factors affecting response effectiveness to an epidemic. In particular, cooperation, trust in others and in institutions -- two key components of social capital -- can help to limit the spread of infectious diseases. Social capital includes a sense of mutual understanding and respect, solidarity, and shared rules (Putnam, 2000). These attributes facilitate helping others and compliance with the containment measures. Mutual understanding and respect, shared rules, and solidarity are crucial components of effective collective action. Ostrom (1991) made it clear that trust helps individuals overcome private incentives in order to cooperate. Experimental evidence also suggests that believing most others will cooperate encourages individuals to do the same (Fischbacher et al. 2001, Shinada and Yamagishi 2007).

Available evidence supports the claim that countries with high social capital fared better during the pandemic than others. Figure 12 presents the correlation between the share of people with high trust in government (as measured in 2016 using European Quality of Life Survey) and the rate of change in new contagions during the first wave of 2020. The correlation indicates that countries in which people trust their government more (on the x-axis) are also countries where new infections declined faster (on the y-axis). This correlation is robust to countries in outlying positions, such as Finland (FIN).

*Figure 12. Confidence in government correlates with the rate of change in new cases*

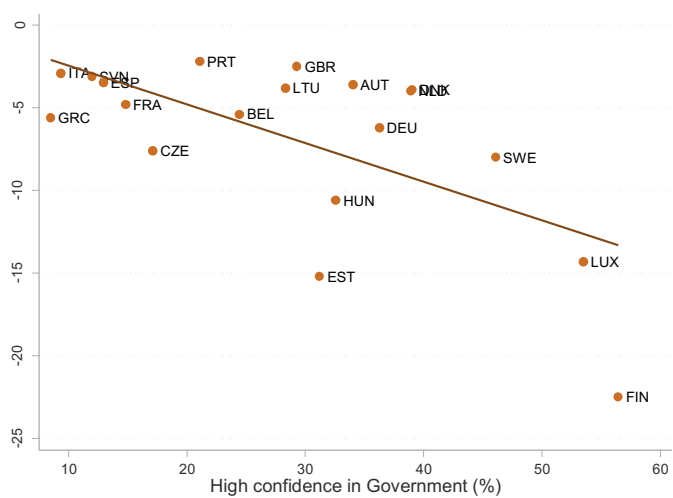

*Note: high confidence in government is defined as people who declared a score higher or equal to 7 on a scale from 1 (not at all) to 10 (a great deal). The threshold was chosen to isolate the group of people that have high trust in government from those choosing intermediate categories such as 5 or 6.*
*Source: own elaboration of Hume Foundation data (May 2020), and European Quality of Life Survey (2016).*

A similar relationship holds if we substitute the rate of change in new infections with mortality (total deaths per one million people). In this case, the correlation coefficient remains negative, although the statistical significance is weak. Bartolini and colleagues (2020) examined this relationship in more detail accounting for the role of various confounders. They created an index of confidence based on people's declared trust in others and in various institutions[6], and they controlled for countries' GDP per capita, inequality, frequency of meeting others, health conditions of the population, number of beds in intensive care units, as well as the number of deaths and government response stringency (before the lockdown). Their results indicate that countries with high trust faced the pandemic faster and with less fatalities: the index of confidence correlates negatively with the rate of change in infections, fewer new cases, and lower mortality (not statistically significant). Bartscher and colleagues (2020) reached similar conclusions using regional data from a sample of seven European countries. In particular, they measured social capital as electoral turnout in the 2019 European elections, but did not include any control variables. They found that areas with high social capital registered between 12% and 32% fewer COVID-19 cases from mid-March until mid-May 2020.

---

[6] The list of institutions include: government, parliament, local authorities, police, press, and judicial system. The data are sourced from the European Quality of Life Survey of 2016.

In sum, there are various reasons to expect that social capital enhances the effectiveness of countermeasures, and available evidence suggests social capital, specifically trust in others and institutions, enhances the efficacy of containment policies. This view is also supported by anecdotal evidence (Pitas and Ehmer, 2020). What is probably less known, however, is that modern societies are organized to exchange social capital for economic growth. This is the prediction of defensive growth models (Bartolini, 2019).

## 5. Defensive growth

Environmental degradation represents just one of the negative externalities associated with mismanaged economic growth. Although such externalities are well known (admittedly agreed upon to varying degrees), the classical representation of economic growth suggests it is always beneficial, as a cake from which everyone gets larger and larger slices. In contrast, defensive growth theory describes a growth that occurs in a vicious cycle, poisoning the cake as it grows. Economic bads (negative externalities) lead to increased consumption. For instance, an increase in sugar production and consumption (due in part to increased prevalence of industrialized food production) may hinder health and increase the demand for pharmaceuticals. Obesity, high cholesterol, and diabetes drive the consumption of anti-cholesterol pills and insulin, and through this channel, generate growth. From this point of view economic growth as a measure of progress loses its appeal.

Defensive growth recognizes that economic growth can be a double-edged sword: negative externalities contribute to economic growth, and additional growth contributes to yet more negative externalities (Bartolini and Bonatti 2003; 2008a; Antoci and Bartolini 2004). The theory assumes that money offers a defense – real or illusory – against the erosion of non-market resources, such as social connections and a pristine natural environment. People's attempt to compensate, or defend their well-being, expands the demand for goods and services, thus fueling consumption and further expanding market activity. Such a growth process entails a substitution process in which market goods and services progressively replace declining non-market sources of well-being.

Marketing has played a significant role in getting the vicious cycle started. In the early 20[th] century, the foundations for marketing as an applied science were present, and by the 1970s to 1980s, the industry became more scientific (Clow and James 2014). They applied insights from a considerable amount of new research to no longer solely advertise individual goods, but to sell lifestyles – as the adage goes, "I shop, therefore I am" – and reach individuals at younger and younger ages, to capture customers for life (Schor 2004). The marketing industry has been tremendously successful and its research advanced our scientific understanding. Unfortunately for us, this created a society dependent on consumption, and driven by materialistic values.

In the happiness literature we know people compare themselves to others; for instance, greater personal income is related to greater happiness, while greater income of others reduces happiness (Clark et al. 2008). The negative relation with others' income is part of a broader phenomenon referred to as social comparison. More recently, the positive relation of personal income was found to be driven by consumption, but not just any consumption, consumption that is easy to compare with others and positional in nature (Wu 2020). Social comparison helps to explain why economic growth has no impact on happiness in the long run (Easterlin 1974; Easterlin and O'Connor 2021). The problem is that positional consumption is a zero-sum game, for there to be winners, there must be an equal number of losers. When happiness depends more on positional consumption than absolute consumption (valued independently of others' consumption), we cannot expect lasting gains from growth. What is worse, individuals still strive for position, to keep up with the Joneses, which leads them to work and consume more. Perversely, this generates economic growth, while ultimately, individuals end up with a house full of appliances and little time to enjoy them or spend with other people.

Together, the rise of materialism and income inequality contribute to social comparisons, which puts pressure on people to make money and to consume, thereby changing values, increasing working time, and limiting the opportunities to establish meaningful social relationships. This process degrades communication and trust, while promoting loneliness and isolation. Indeed, loneliness was described as an epidemic in the United States (McGregor 2017), and the United Kingdom has a Minister of Loneliness (Walker 2018).

Degradation of the natural environment likewise contributes to the cycle. As discussed above, human activity has led to a significant acceleration in biodiversity loss. Global Warming is similar. Historical growth was fueled to a large extent by a reduction in the cost of energy arising from the discovery of hydrocarbons, the burning of which contribute to the greenhouse gas emissions leading to Global Warming. More immediately, few people want to, nor should, swim downstream from an industrial facility or near a large port. Consequently, people invest in private pools and go on vacations. Perversely, clean-up efforts to mitigate environmental damage contribute GDP growth.

In sum, the degradation of social and natural environments reduces well-being and people seek remedies to compensate for their loss. The market, with the help of the advertising industry, offers quick and private remedies to every problem, including poor relationships. People who are looking for social interactions, but have limited time, can book a date with a Moomin -- one of the characters of a famous Finnish series of books and comic strips; if they lack the warmth of a pet, they can buy an android pet, without having to take care of them; if people like to sing, they can rent an individual karaoke booth to sing to themselves. The truth, of course, is that goods do not love, they are as lifeless and inert as they have ever been. This disillusion feeds consumer frustration and sets the ground for endless consumption. Not by chance, the economy of loneliness and fear is a booming sector in many developed and developing countries, yet people are no less lonely than before. Likewise, if their drinking water is polluted, people can install filters to purify it; if the place where they live is too noisy, they can install triple glass windows and insulate their houses with the latest product available on the market. In all these cases, people adopt private solutions to address common problems. The tragedy is that the sum of the individual efforts ends up worsening the living conditions of all. Individuals attempt to compensate for environmental and social degradation, thereby fueling further growth. Economic growth can therefore be the result of a self-perpetuating, vicious cycle in which economic expansion is the cause and consequence of its harmful effects on the environment, society, and ultimately, well-being.

There are many additional examples of people's efforts to compensate for fewer social resources. Families with insufficient time can hire care-givers; if people are lonely, their friends are too far away, or the city is too dangerous to be out at night, they can purchase home entertainment systems. Work environments characterized by distrust can be extremely difficult, time-consuming, and nerve-wracking. To compensate, companies pour considerable resources into observing and incentivizing employees, as well as programs to cultivate a positive social environment. Think of the many solutions available on the market to control employees' work activity, or the legal expenses to write sophisticated contracts to prevent or discourage free riding and moral hazard. In all these cases, people adopt private solutions to common problems, which is a clear example of coordination failure.

The root cause of this vicious cycle is a fundamental lack of cooperation and coordination, which pushed people to seek private solutions because social action was impossible. In previous examples, people would be better off if they cooperated and adopted common solutions. However, if economic growth erodes social relations – including trust in others and in institutions – the possibility to cooperate decreases with time. In the United States, the share of people trusting others and Congress has been steadily declining: in mid-70s nearly 20% of Americans declared they trusted Congress, and 45% trusted others. Forty years later, the share of Americans trusting Congress amounted to slightly more than 5%, whereas the share of people trusting others declined to nearly 30% (based on data from the General Social Survey, a large scale, nationally representative survey of Americans). Other developed and developing countries have similar experiences, such as the United Kingdom and China.

If people do not trust others and their institutions, they will lose confidence in the efficacy of collective action, and they will look for private solutions to compensate for the depletion of non-market resources. Thus, defensive growth creates its own fortune by eroding non-market resources and by changing common problems into private issues. Economic growth, as well as unsustainability, therefore, results from the sum of many private answers to common problems. Poor quality of life, the antithesis of progress itself, is the corollary of an economic growth that is driven by defensive needs.

In short, defensive growth theory predicts: environmental degradation, the erosion of social relations (in both developed and developing countries), as well as long working hours, stagnating well-being, consumerism, and declining trust in others and in institutions (Bartolini et al., 2014). These predictions have been the subject of empirical scrutiny in recent years. In particular, available studies explored whether money and social relationships are substitutes, whether economic growth can erode social capital and impede well-being, and whether low social capital predisposes materialistic attitudes, growing consumption, and long working hours (for a review of these studies, see Sarracino and Mikucka, 2019).

When growth is defensive, the declining quality of the environment and relationships, as well as high workload, offset the positive effects of income growth and impede well-being. This contributes another explanation to why greater economic prosperity may not be associated with greater well-being, part of the oft-cited Easterlin Paradox (Easterlin 1974; Easterlin and O'Connor 2021).

# 6. Neo-humanism: call to action

COVID-19 illustrates how much our ability to survive depends on cooperation. Epidemics are more immediate and tangible than other common challenges -- such as climate change -- that, on the contrary, have less apparent and direct consequences. Many environmental challenges seemed too uncertain and far away to be taken seriously. COVID-19 has changed this by putting people's health at stake in a remarkably short time. The good news is that COVID-19 captured serious attention and gave us the opportunity to rethink the world in which we live.

Once the emergency is over, it will be the time to change the way modern societies are organized, to finally make them compatible with people's needs: positive inter- and intra- personal relationships and with the natural environment. Indeed, there are already initiatives in place to "Build Back Better"[7]. Broadly, these initiatives hope to use stimulus money to target social and environmentally inclusive ends. Alternatives that ignore our current environmental challenges promise grim futures.

Changing society is not easy of course. It requires a deep reform in organization, believing in the importance of social relationships and cooperation, and abandoning the idea of economic growth at any cost. Economic growth, if it is defensive, may be more of indicator of backward rather than forward progress. As long as the emphasis in economics is to maximize profits, it cannot be trusted to develop an agenda to promote environmental and social sustainability.

The words of the Nobel Peace Prize winner, Muhammad Yunus, are particularly insightful: "First and foremost, we have to agree that the economy is a means to facilitate us to reach the goals set by us. It should not behave like a death trap designed by some divine power to punish us. We should not forget for a moment that it is a tool made by us. We must keep on designing and redesigning it until we arrive at the highest collective happiness. If at any point, we feel that it is not taking us where we want to go, we must immediately know that there is something wrong and fix it. […] It is all about building the right hardware and the right software. The power is in us. When human beings set their minds on something, they get it done. Nothing is impossible."

---

[7] See for instance (Hamann 2020) and references therein.

It is time to put humans, and their well-being, at the center of decision making. The good news is that the studies on quality of life have reached a considerable degree of maturity, enough to inform the development of a new social and economic organization, as part of the neo-humanism movement. Note, we emphasize humans, but neo-humanism includes the natural environment as well, for both intrinsic and extrinsic reasons.

## 6.1 What is neo-humanism

Neo-humanism is a movement to put humankind back at the center of societies' attention. It is grounded on recognizing that GDP is not an indicator of well-being and that its preeminent position in policy-making has diverted attention from important aspects of people's lives, such as their relationships with others and the environment. It recognizes that the user-friendliness of GDP led to policies that may serve the markets well, but not necessarily humankind or the environment. Indeed, it is difficult to say these policies performed well even in terms of the GDP growth rates of Western countries over the past 40 years (Figure 14). The picture worsen if we consider the social and environmental damage inflicted over this period.

Neo-humanism is a cultural movement that proposes a shift from the "business as usual" status quo. This shift requires "holistic" policies, i.e. policies designed to account for their direct and indirect effects on people's well-being. To clarify, neo-humanism does not argue for de-growth, but refutes the agenda of growth at any cost: societies should grow in a socially and environmentally compatible way. Indeed, economic growth can be compatible with well-being in countries that promote full employment and social safety nets (Easterlin, 2013; Ono and Kristen, 2016), protect social capital (Uhlaner 1989; Helliwell 2003, 2008; Bartolini et al. 2013; Clark et al. 2014), and reduce income inequalities (Oishi and Kesebir, 2015; Mikucka et al., 2017; Sarracino and O'Connor 2021). In these countries the economy might grow slower than elsewhere, but slow or near-zero economic growth is not necessarily a bad sign. On the contrary, it may signal a system that is better organized to support quality of life. We should abandon the common idea that economic growth is always good and introduce a new definition of performance, corresponding to societies' ability to transform resources into quality of life.

*Figure 13. GDP growth rates by decade in a selected sample of countries.*

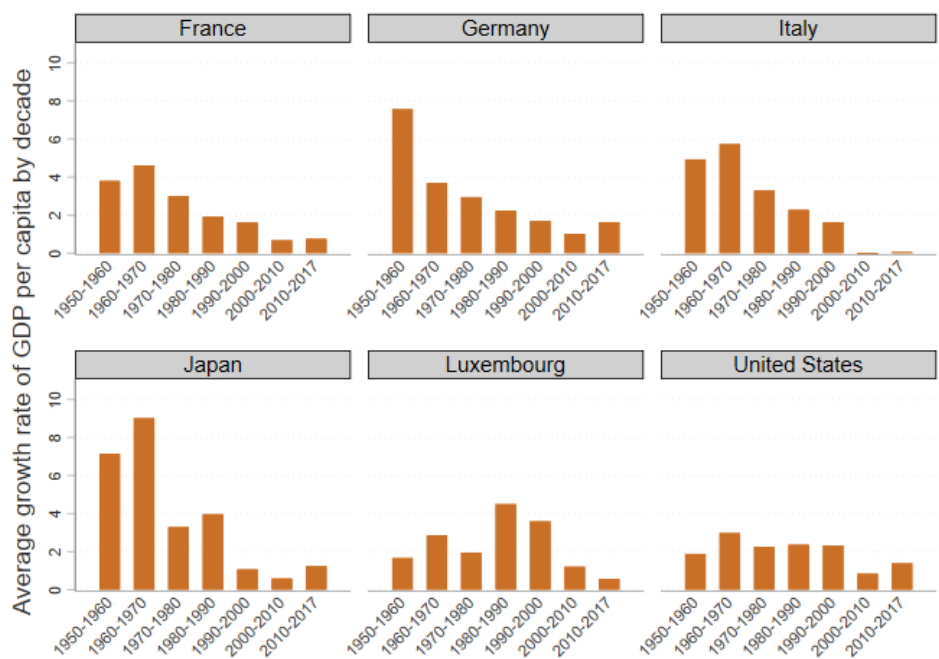

Note: Real GDP at constant 2011 US$, in millions.
Source: Own elaboration of Penn World Tables data (ver. 9.1).

This is by no means the first movement to go Beyond GDP to measure progress or promote change (e.g., Kubiszewski et al. 2013; Fleurbaey 2009). The social indicators movement gained a dedicated journal in 1974, aptly named, *Social Indicators Research* (Sirgy et al. 2006), though the movement really picked up steam in the 2000s. In 2004, the OECD began an agenda to improve measures of progress, culminating in the Sarkozy Commission (Stiglitz et al. 2009), Global Policy Reports (The Global Happiness Council 2018), reports on measuring subjective well-being and advocating nations to do so (OECD 2013). Neo-humanism draws from their legacies; it differs from other religions, philosophies, and world views in its objectives. According to neo-humanism, well-being is not prescribed; it is what people consider as such, grounded on evidence from quantitative analyses of people's own evaluations of their lives.

Easterlin (2019) and Layard (2020) argue to supplant GDP with subjective well-being (life satisfaction) as the preeminent measure of progress, in part because subjective well-being is user-friendly (indeed more relatable than GDP), while alternatives, for instance dashboards of indicators, are less user-friendly and prone to selective reporting by stakeholders. Subjective well-being can be manipulated too (Frey and Stutzer 2010), though there is no present evidence of this.

Neo-humanism does not argue for one measure over another, but for a change of culture, transitioning from material and income-based objectives to more holistic quality of life objectives. Rather than conceive of income as the preeminent measure of the good life, we would learn from the quality-of-life research what factors contribute to greater well-being, both public and private. There are various tools already in place to track quality of life, as individuals, in firms, and communities (large and small).[8] Individuals can learn for themselves what the research implies for them, and policy makers can promote settings for greater well-being.

## 6.2 Changing the cycle, from vicious to virtuous

The literature on quality of life provides a number of insights on how to organize a socially and environmentally sustainable future. The first step is to promote social relationships.

Investing in social relations could break the self-reinforcing defensive growth cycle. As previously explained, individuals compensate for poor and deteriorating social relations with defensive or palliative consumption, which contributes to growth. Promoting social relations addresses the defensive cycle at multiple points. Ample social relations reduce the need to compensate with goods and services, thereby reducing consumption, which frees up working time and reduces the negative externalities associated with excess consumption. In turn, reducing negative externalities puts even less pressure on individuals to compensate. Ample social relations also contribute to trust, in others and institutions, which facilitate the collective action necessary to address negative externalities.

Indeed, recent evidence indicates that income plays a lesser role for the well-being of people with more social relations, implying they are less driven to compensate or defend their well-being than those with poorer social relations (Bartolini et al., 2019). The authors illustrate this finding using Figure 15. The X axis presents the share of people with high social capital in a region, while the Y axis presents the gap in life satisfaction between the rich and poor in that region. The negative and significant slope indicates that regions with high social capital are regions in which income is less associated with subjective well-being.

---

[8] See, for instance, Mappiness (http://www.mappiness.org.uk), OECD's Better Life Index (http://www.oecdbetterlifeindex.org/#/11111111111), or multiple apps made available by What Works Centre for Wellbeing (https://whatworkswellbeing.org/resources/?_sft_resource-type=app-online-tool).

Social relations are also a well-established (Becchetti et al., 2009; Helliwell and Aknin 2018) and lasting component of subjective well-being. While individuals' subjective well-being tends to adapt to numerous life circumstances, adaptation to social relationships is only partial (Clark, 2016). Meaning, that investing in social relations will have a more lasting impact on subjective well-being. And, greater subjective well-being in turn contributes to a virtuous cycle.

*Figure 14. The life satisfaction gap between rich and poor people is smaller in regions with a rich social life.*

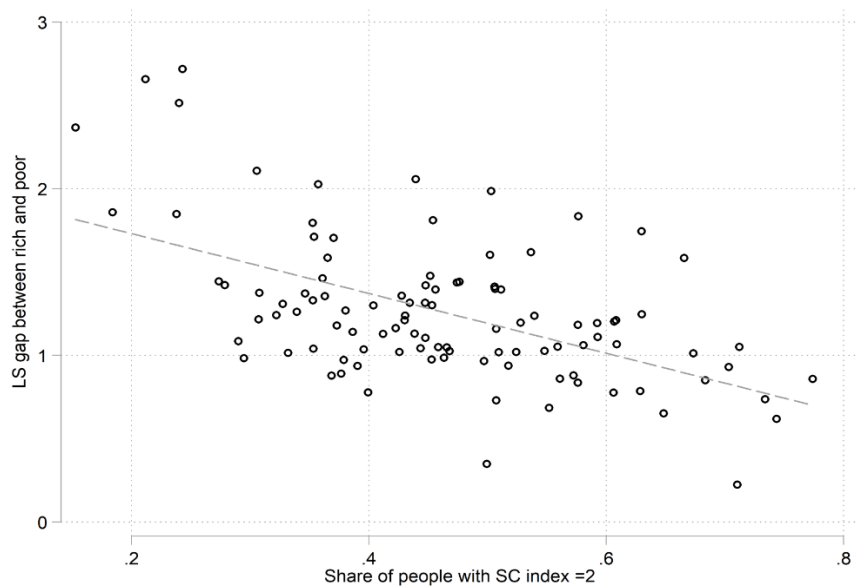

*Note: Social capital is measured as the share of respondents with a social capital index equal to 2. The social capital index has a maximum score of 2 if a person trusts others and meets friends at least once a month. Aggregated data are computed from individual data using sample weights. Data: EU-SILC, 2013. N = 99 European regions.*
*Source: Bartolini et al., 2019.*

There is a rich literature showing that happy people are also more productive (Oswald et al., 2014; Proto et al., 2010); they are more pragmatic, less absent, more cooperative and friendly (Judge et al. 2001); they change jobs less frequently, and are more accurate and willing to help others than less happy people (Spector, 1997). Available evidence also indicates that happier people are more engaged at work, earn more money, and have better relationships with colleagues and customers (George and Brief, 1992; Spector, 1997; Wright and Cropanzano, 2000), and are less likely to be unemployed (O'Connor 2020b). Each aspect is related to productivity and job performance. In particular, DiMaria and colleagues (2020) computed that an increase of one unit in life satisfaction in a country such as Germany or France contributes to productivity gains that are comparable to nearly 80 working hours per year. In other words, a unit increase in life satisfaction (on a scale from one to ten) would allow people to work nearly two working weeks less while leaving the output unchanged. This means that by increasing people's well-being, it is possible to free resources that can be used to implement projects for well-being, for instance to: cultivate personal interests, dedicate time to others; build social relationships; and contribute to collective action. What is more, subjective well-being has numerous positive benefits in other domains as well, including social capital (Guven 2011) and health (Tay et al. 2015). See De Neve et al. (2013) or Piekałkiewicz (2017) for summaries.

The evidence suggests people are more open to a change than one might think. The desire to over consume is not strictly rooted in people's greed, as is commonly believed, but is a consequence of the features of the socio-economic organization as described above. In fact, people seem to care substantively for the future environment. Those who expect the distant future to be bleak, are less satisfied with their lives (Bartolini and Sarracino 2018). Importantly, in this study the future is distant enough to not involve the respondents or their direct descendants; meaning, the results imply individuals reveal to be intrinsically motivated to save the environment, if only they could; if others could be trusted, they would prefer to coordinate their actions to address negative externalities and reduce palliative consumption. Indeed, the negative association between life satisfaction and bleak-future-expectations is relatively large, comparable in magnitude to becoming unemployed or getting married.

Promoting social relations could interrupt the defensive-vicious cycle and instill a virtuous one. The pursuit of well-being, not income, should decrease consumption, thereby limiting negative externalities, benefit the environment, and create better conditions for cooperation and more prosperous societies. Increased well-being also contributes to productivity, thus benefiting economic growth, but a growth that is driven by creativity, not palliative consumption; a growth that is decoupled from people's ability to enjoy a good life; perhaps a slower growth, but one that is better suited to fit people's needs.

## 7. Conclusion

Economic growth does not necessarily improve people's lives and when prioritized and mismanaged, it contributes negatively. It is now more than 10 years since international institutions, backed by authoritative thinkers, have called for going "beyond GDP". What would such a world look like? And how do we get there? In this article, we propose neo-humanism as a reference to promote a future where well-being is decoupled from economic growth.

Neo-humanism is a cultural movement to put humankind back at the center of decision-making. Just like the humanists in the early fifteenth century aimed to rediscover the authentic messages of classic philosophers for the sake of a new, egalitarian, and independent society, neo-humanism sets to re-discover the foundations of what makes a life worth living, and proposes to re-organize modern societies accordingly.

Neo-humanism is grounded on the idea that the preeminence of GDP in policy, social discourse, and media has diverted attention from other important aspects of people's lives, such as their relationship with others and the environment. Neo-humanism recognizes that the user-friendliness of GDP led to unidirectional policies that may serve the markets well, but not necessarily humankind or the environment. The erosion of social and natural environments -- that is widely recognized in academic environments, and probably contributed to the onset and uncoordinated response to the pandemic -- are the result of such myopia.

Neo-humanism proposes a change of culture informed by self-reported measures of well-being, i.e. a spontaneous, non-mediated, and democratic assessment of individuals' lives as a whole. The interdisciplinary field on quality of life applies qualitative and quantitative methods to the analysis of these reports and provides a number of insights concerning the good life. By organizing the evidence from various studies and different perspectives, we sketch how to shift from income as the preeminent measure to promoting well-being, i.e. how to put humans back at the center of decision making. The role of policy makers would be, therefore, to create the conditions for people to flourish and lead the lives they wish. The studies summarized in this work suggest that this would contribute to a socially and environmentally sustainable future.

It is first important to understand when economic growth fails to improve human well-being. According to defensive growth theory, individualistic societies privilege private solutions to common problems. However, the sum of individuals' actions worsens the initial problem, thus generating a vicious cycle whereby the more people are concerned by a common problem, the more their reactions worsen the problem. This cycle leads to societies in which the importance of money and working hours increases, along with loneliness, consumption, environmental degradation, and unhappiness. The good news is that there is an alternative. Recent studies show that economic growth may contribute to increasing well-being when it is accompanied by generous welfare schemes, good social relations, and low income inequality; in other words, when it takes places in an inclusive society.

This body of work indicates that it is possible to replace the defensive growth cycle with a virtuous one by adopting policies for well-being, such as promoting mutual trust and cooperation, two key components of social capital. In happier societies, people's need of defensive consumption is low, which benefits the environment, reduces people's dependence on money to lead a good life, and contributes to cooperating and prosperous societies. What is more, the idea that promoting happiness would reduce incentives to work and put societies on snooze is incorrect. In fact, happier people are more productive. Thus, greater happiness contributes to economic growth, but a growth that is driven by creativity, not defensive consumption; perhaps a slower growth, but one that is better suited to fit people's needs.

COVID-19 illustrates the limits of economic growth as a measure of well-being, and the importance of protecting common goods, such as social and natural environments, in individualistic societies. It also gave us the possibility to re-think the world in which we want to live. COVID-19 affects everyone, some more than others, it is true. But a world in which we work together may have prevented COVID-19, as well as the 2008 economic crisis. Even those that are adept at private solutions, cannot diversify against such systemic risk. Neo-humanism seeks a world in which the well-being of people comes before the well-being of markets; a world in which promoting cooperation and social relations is the starting point for better lives and a peaceful and respectful coexistence with other species on Earth.